\documentclass[aps,prl,reprint,groupedaddress]{revtex4-1}

\usepackage{amsmath}
\usepackage{hyperref}
\usepackage{epsfig}
\usepackage{graphicx}
\usepackage{subfigure}
\usepackage{latexsym}
\usepackage{color}
\usepackage{fullpage}
\usepackage{epstopdf}
\usepackage{dcolumn}
\usepackage{bm}
\usepackage{ulem}
\usepackage{units}

\newcommand{\Ioan}[1]{\textcolor{black}{#1}}

\begin{document}

\title{Ground-State Candidate for the Dipolar Kagome Ising Antiferromagnet}

\author{I. A. Chioar$^{1,2}$, N. Rougemaille$^{1,2}$ and B. Canals$^{1,2}$} 

\address{$^1$CNRS, Inst NEEL, F-38000 Grenoble, France\\$^2$ Univ. Grenoble Alpes, Inst NEEL, F-38000 Grenoble, France}

\date{\today}

\begin{abstract}
We have investigated the low-temperature thermodynamic properties of the dipolar kagome Ising antiferromagnet using at-equilibrium Monte Carlo simulations, in the quest for the ground-state manifold. In spite of the limitations of a single spin-flip approach, we managed to identify certain ordering patterns in the low-temperature regime and we propose a candidate for this unknown state. This novel configuration presents some intriguing features and passes several test-criteria, making it a very likely choice for the dipolar long-range order of this kagome Ising antiferromagnet.	
\end{abstract}

%\pacs{75.10.Hk, 75.50.Lk, 75.70.Cn, 75.60.Jk}

\maketitle

\section{I. Introduction}

Dipolar long-range interactions can often lead to unconventional arrangements of magnetic moments, particularly when their interplay with the spatial distribution of these moments results in geometrical frustration. One typical example can be found in spin ice pyrochlores, such as Ho$_2$Ti$_2$O$_7$ \cite{harris_geometrical_1997} and Dy$_2$Ti$_2$O$_7$ \cite{ramirez_zero-point_1999}, where the observed ice-like physics has a deep dipolar root \cite{den_hertog_dipolar_2000}. Artificial realizations that mimic the frustration-induced effects encountered in such condensed matter compounds have recently attracted much interest and have generally taken the form of lithographically-patterned arrays of magnetic nano-islands, bearing the name of artificial spin ices \cite{wang_artificial_2006,tanaka_magnetic_2006}. The magnetostatic framework of these arrays, the almost infinite freedom in design and the possibility to locally probe each magnetic component have enabled the exploration of a wide \Ioan{range of intriguing phenomena over the past few years} \cite{nisoli_colloquium:_2013,heyderman_artificial_2013-1}. 

\begin{figure}
	\includegraphics[width=1.0\linewidth]{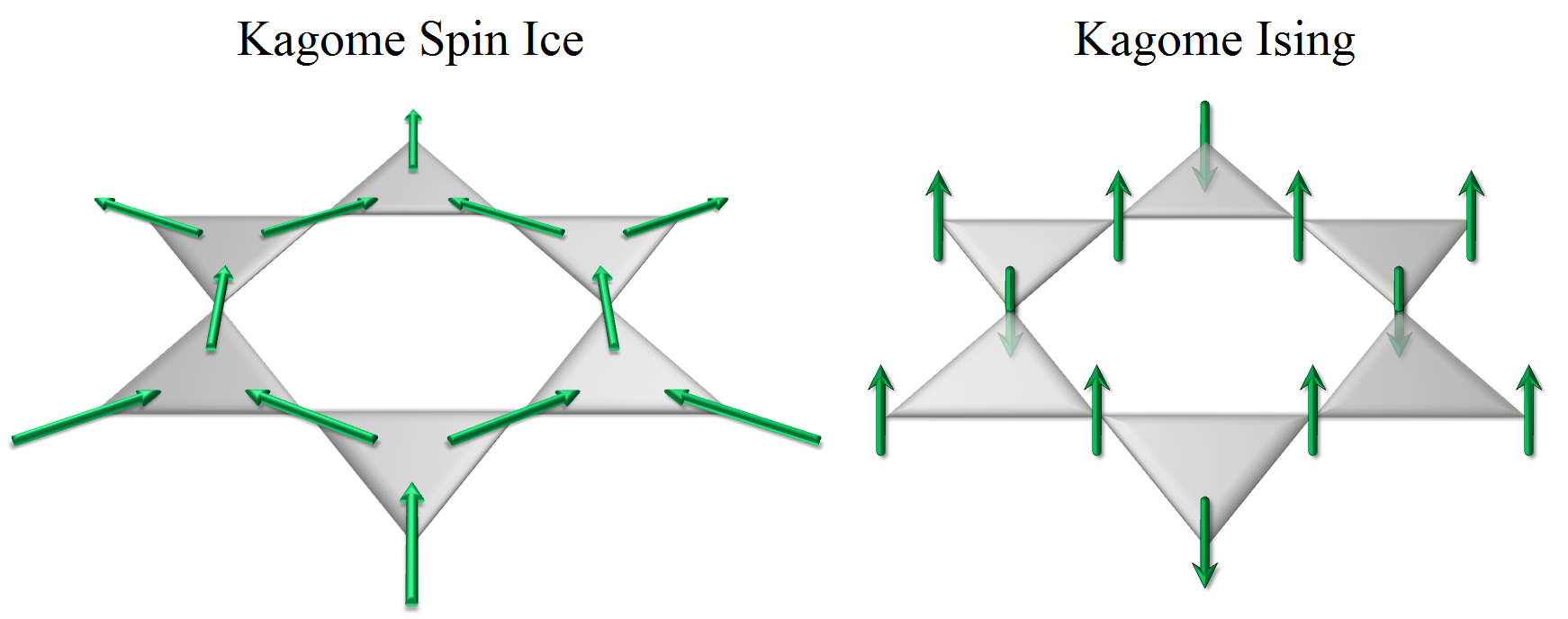}
	\caption{\label{Figure1}
		(Color online) The kagome lattice is a triangular network of corner-sharing triangles. Two particular arrays have been fabricated so far in the framework of artificial spin ice:  the kagome spin ice, with the Ising-like spins lying in the network plane, and the kagome Ising, with spins pointing along the vertical axis.}
\end{figure}

\Ioan{Several artificial frustrated systems have been recently studied}, with a particular attention given to the square and kagome geometries. The kagome network is one of the most frustrated two-dimensional lattices and displays some exotic thermodynamic features. In particular, dipolar kagome spin ice, with magnetic moments lying along the bisectors of each triangle (see Figure~\ref{Figure1}), presents four distinctive temperature regimes and two critical phenomena before it achieves long-range spin order \cite{moller_magnetic_2009,chern_two-stage_2011}. More specifically, this system first exhibits a cross-over from the paramagnetic phase into a spin ice manifold, where each triangle respects the so-called kagome ice rule, requiring the existence of a minority spin per triangle. This local constrain minimizes nearest-neighbor interactions and, in a short-range interaction picture, yields a macroscopically degenerated manifold \cite{syozi_statistics_1951,kano_antiferromagnetism._1953} characterized by a cooperative paramagnetic regime \cite{villain_insulating_1979} and exponentially-decaying pairwise spin correlations. Longer range couplings continue to further correlate the system as the temperature is lowered, but prior to achieving long-range order, dipolar kagome spin ice first passes through an intermediate phase in which \Ioan{spin order and disorder coexist} \cite{brooks-bartlett_magnetic-moment_2014}. This so-called spin ice 2 state is also regarded as an algebraic spin liquid that sits upon a magnetic charge crystal, given the dumbbell charge description \cite{castelnovo_magnetic_2008,moller_magnetic_2009,chern_two-stage_2011}. Driven by these exotic properties, artificial realizations of dipolar kagome spin ice have been intensively studied, first by employing field demagnetization protocols \cite{qi_direct_2008,li_comparing_2010-1,rougemaille_artificial_2011} and, more recently, by using thermally-active arrays, which have facilitated the local access of the exotic spin ice 2 manifold \cite{zhang_crystallites_2013,montaigne_size_2014,chioar_kinetic_2014,drisko_fepd_3_2015}. 

\begin{figure*}
	\includegraphics[width=1.0\linewidth]{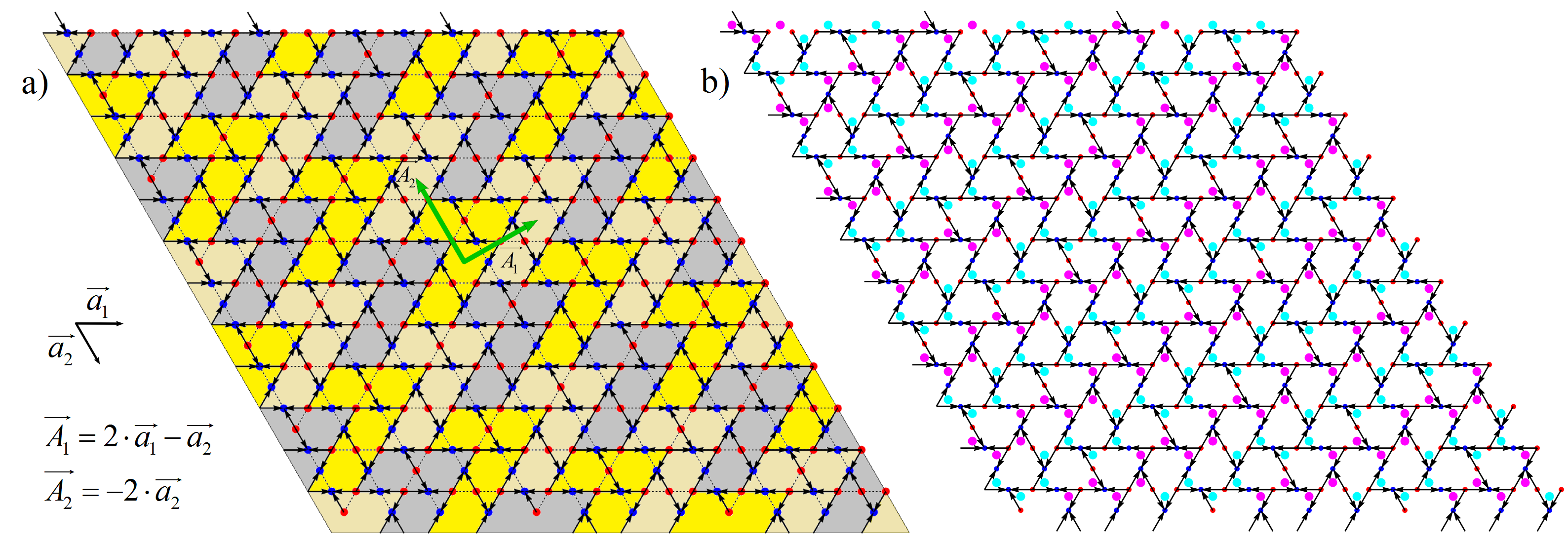}
	\caption{\label{Figure2}
		(Color online) The 7-shaped phase - a candidate for the ground-state of the dipolar kagome Ising network. (a) The spin configuration is a crystal of 7-shaped unit cells with a rectangular basis ($\vec{A_1}$,$\vec{A_2}$), which can be made commensurable with the Bravais triangular basis ($\vec{a_1}$,$\vec{a_2}$) of the original kagome network. Each unit cell contains (a) 12 spins, represented here by red and blue dots corresponding to the \textit{up} and \textit{down} states respectively, and (b), if the dumbbell charge description is considered, 8 vertex magnetic charges, represented by magenta and cyan dots for the $+1$ and $-1$ states, respectively. \Ioan{The arrows that reveal the 7-shaped unit cells form} the so-called \textit{arrow picture}, which maps the local stray field lines from all \textit{up} spin states to all their nearest-neighboring \textit{down} spin states. Given the rotational symmetry of the unit cell and time-reversal symmetry, this phase is sixfold degenerate.}
\end{figure*}

Nevertheless, another type of kagome network has been fabricated \cite{zhang_perpendicular_2012,chioar_nonuniversality_2014}, with magnetic moments pointing perpendicular to the lattice plane (see Figure~\ref{Figure1}). This so-called kagome Ising system presents only antiferromagnetic pairwise couplings, contrary to its in-plane counter-part, which displays a mixture of ferro- and antiferromagnetic interactions, depending on the relative orientation of the considered pair of spins. The two kagome networks show identical thermodynamic behavior if governed solely by nearest-neighbor couplings, but dipolar long-range contributions set them apart and each one of them has its own story to tell. In other words, these longer range couplings enrich the palette of exotic magnetic phases that can potentially be achieved within the framework of artificial spin ices and a simple change in the spin orientation can yield completely new spin textures even when the lattice topology is preserved. However, while the kagome spin ice system has been intensively explored, the low-temperature properties of this dipolar kagome Ising antiferromagnet remain mostly unaddressed, both theoretically and experimentally, and a main challenge is to investigate the potential formation of long-range spin order. 

By employing Monte Carlo simulations, we have explored the at-equilibrium thermodynamic features of this spin model, particularly in the low-temperature regime. Although our Monte Carlo approach does not manage to simulate the full recovery of the magnetic degrees of freedom, we believe that this system ultimately achieves long-range spin order and we provide a candidate for its dipolar long-range ground-state. This complex and unexpected configuration has a 12-spin magnetic unit cell that forms a rectangular crystal, commensurable with the underlying triangular Bravais lattice, as reproduced in Figure~\ref{Figure2}. Interestingly, by taking the dumbbell description and thus replacing the spins with classical magnetic charges that can ultimately be merged to defined an effective magnetic charge per triangle \cite{chioar_nonuniversality_2014}, this configuration reveals an alternating sequence of charged stripes, in contrast with the antiferromagnetic charge order of the dipolar kagome spin ice \cite{moller_magnetic_2009,chern_two-stage_2011}. Although some studies have been previously performed on the kagome Ising antiferromagnet with second and third order couplings \cite{azaria_coexistence_1987,wolf_ising_1988,takagi_magnetic_1993,wills_model_2002}, to the best of our knowledge, this state has not been reported before. It also satisfies several criteria, making it a very likely choice for the lowest-energy state of the dipolar kagome Ising antiferromagnet. 

The remainder of this paper is organized into three main sections. We shall first describe the model employed along with the parameters of our Monte Carlo simulations. Afterwards, a specific section is dedicated to the testing of this candidate as a ground-state, a configuration to which we shall refer to as the 7-shaped phase. Furthermore, a discussion will follow focusing on some of the intriguing features that this state displays. Lastly, conclusions are provided along with potential routes for further investigation.

\section{II. The Model and the Simulations}

We have performed Monte Carlo simulations for kagome Ising networks with $L\times L \times 3$ lattice sites ($L$ ranging from $12$ to $36$) using the dipolar spin ice Hamiltonian \cite{den_hertog_dipolar_2000}. Given the Ising nature of the spins and the fact that they all share the same quantization axis ($\vec{e_z}$), the Hamiltonian can be brought to a more compact, scalar form:
\begin{align}
H = J_0\sum_{<u<v>} \sigma_u \cdot \sigma_v + D \cdot \sum_{(u<v)} \frac{\sigma_u \cdot \sigma_v}{r_{uv}^{3}}
\label{eq:eq1}
\end{align}
where $\sigma_u$ and $\sigma_v$ represent the Ising states ($\vec{S}_u = \sigma_u \cdot \vec{e}_z$ and $\sigma_u = \pm1$), $r_{uv}$ stands for their relative position, $D$ is the dipolar constant and $J_0$ is an additional coupling that corrects the interaction between nearest-neighbors, should it deviate from the dipolar approximation \cite{Note_Micromagnetic_Deviation}. This allows us to define the nearest-neighbor coupling as $J_{NN} = J_0 + D/r_{NN}^3 = 3/2 \cdot D/r_{NN}^3$. Periodic boundary conditions have been implemented and the pairwise couplings are summed up within a maximum radius disk such that a spin does not interact with one of its images \cite{Note_Cut_off}. Starting from a paramagnetic regime ($T/J_{NN} = 100$), the temperature is gradually dropped down. A Metropolis single-spin flip algorithm has been employed \cite{metropolis_equation_1953} and, for each temperature value, we used $10^4$ modified Monte Carlo steps (MMCS) for thermalization followed by $10^4$ MMCS for sampling decorrelated spin configurations and computing relevant thermodynamic quantities (see note \cite{Note_Modified_Monte_Carlo_Steps} for more details on MMCS).

After the system reaches the spin ice phase and all triangles respect the kagome ice rule, longer range contributions further correlate the system. The single-spin flip dynamics still manages to ergodically map the phase space down to a deep spin ice temperature, but it ultimately suffers from a critical slowing down effect for $T/J_{NN} \cong 0.03$, at a point where the system seems to be experiencing a phase transition. A similar phenomenon occurs for dipolar kagome spin ice at the onset of the spin ice 2 phase, but this algorithmic inconvenience can be overcome by including collective spin flips \cite{wolff_collective_1989} in the form of closed-loops \cite{melko_long-range_2001,moller_magnetic_2009,chern_two-stage_2011}. These loop updates ensure that the kagome ice rules are not violated while they also preserve an already-established antiferromagnetic charge order. The former argument has motivated us to apply this collective spin dynamics to the kagome Ising system as well, where the loops become alternating, closed-chains of spins. While this addition enhances the exploration of the warm spin ice regime, these collective spin flips are increasingly rejected at lower temperatures and this particular dynamics ultimately freezes at roughly the same point as the single spin-flips do. Nevertheless, the thermodynamic behavior sampled so far proves to be useful for testing the validity of the 7-shaped phase as a potential ground-state configuration.

\section{III. Testing the \textit{7-shaped} Phase}

\begin{figure*}
	\includegraphics[width=1.0\linewidth]{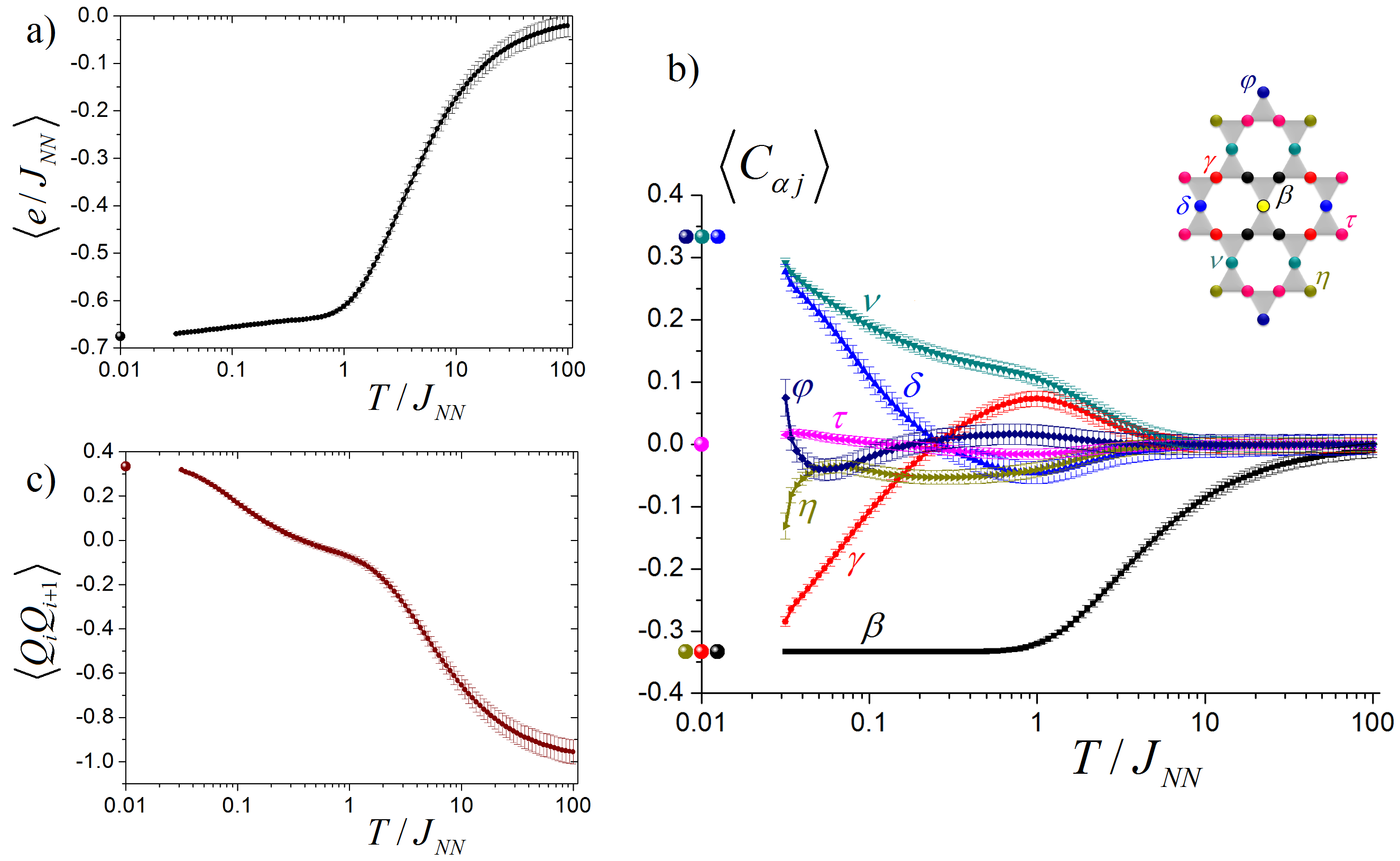}
	\caption{\label{Figure3}
		(Color online) The temperature dependencies of the average values of different thermodynamic quantities computed for a dipolar kagome Ising network with $36 \times 36 \times 3$ lattice sites within the simulated temperature window, prior to the critical slowing down of the single-spin flip dynamics. In all three cases, the error bars correspond to the standard deviation of the distribution of values sampled for each Monte Carlo temperature and the colored dots highlight the values extracted from the candidate state. (a) After minimizing nearest-neighbor interactions, the average energy continues to decrease steadily, reaching a value of $\langle e_{LT}/J_{NN} \rangle = -0.6702$ for the lowest simulated temperature, slightly higher than the 7-shaped phase one, $e_{7B}/J_{NN} = -0.6754$. (b) The evolution of the first seven pairwise spin correlators, as defined in the inset. All plots aim at the values computed from the 7-shaped phase, thus reflecting the system's apparent desire to reach it. With the exception of $C_{\alpha \tau}$, all values tend to $\pm 1/3$. In fact, the $\alpha\beta$, $\alpha\gamma$ and $\alpha\eta$ pairs form kagome networks that fully respect the kagome ice rule. Therefore, in this candidate state, the short-range constrain is applied to certain longer range couplings as well. (c) The nearest-neighbor charge correlator displays a monotonous behavior, continuously increasing and apparently aiming at the value $+1/3$, compatible once again with the proposed configuration.}
\end{figure*}

Firstly, for this candidate to truly be a ground-state, its energy has to be the lowest possible within the dipolar framework. In fact, the energy/spin of this configuration is $e/J_{NN} = -0.6754$ and it is also lower than all the other values sampled throughout the entire simulation (see Figure~\ref{Figure3}.a). Notice the relatively small difference between this value and the energy of the short-ranged ground-state, which, for the current parameters, would be $e_{SR}/J_{NN} = -2/3$. While this slight difference might conceals the importance of longer range coupling terms, their impact is highlighted by the temperature evolutions of the network averages of the pairwise spin correlators, $C_{\alpha j}(T) = \langle\sigma_{\alpha} \sigma_j \rangle(T)$, which can also serve as test-agents for the 7-shaped state.

The development of the first seven pairwise spin correlators, as defined by Wills \textit{et al.} \cite{wills_model_2002}, is given in Figure~\ref{Figure3}.b, along with the values extracted from our ground-state candidate. Notice how the latter play the role of target values for the evolution of the former, thus emphasizing the compatibility of the 7-shaped state with the overall thermodynamic behavior. The same feature can be observed if the kagome spin network is effectively replaced with a hexagonal array of magnetic charges and the nearest-neighbor charge correlator, $Q_iQ_{i+1}$, is computed. As previously reported \cite{chioar_nonuniversality_2014}, the kagome Ising charge correlator becomes positive deep within the spin ice phase and appears to evolve towards the value of $+1/3$ (see Figure~\ref{Figure3}.c). This would imply that, on average, each vertex charge is surrounded by two nearest-neighbors of the same sign and one with the opposite. This is exactly the case in the 7-shaped phase, where each local charge correlator is equal to $+1/3$, given the formation of winding charged stripes. While it may be argued that the correlators vary significantly during a phase transition, which the system appears to experience at the point where the simulation freezes, these changes are made with the goal of energy minimization. Therefore, the first three tests performed so far (Figure~\ref{Figure3}.a-c) are not only passed successfully, but they also support each other. 

\begin{figure*}
	\includegraphics[width=0.8\linewidth]{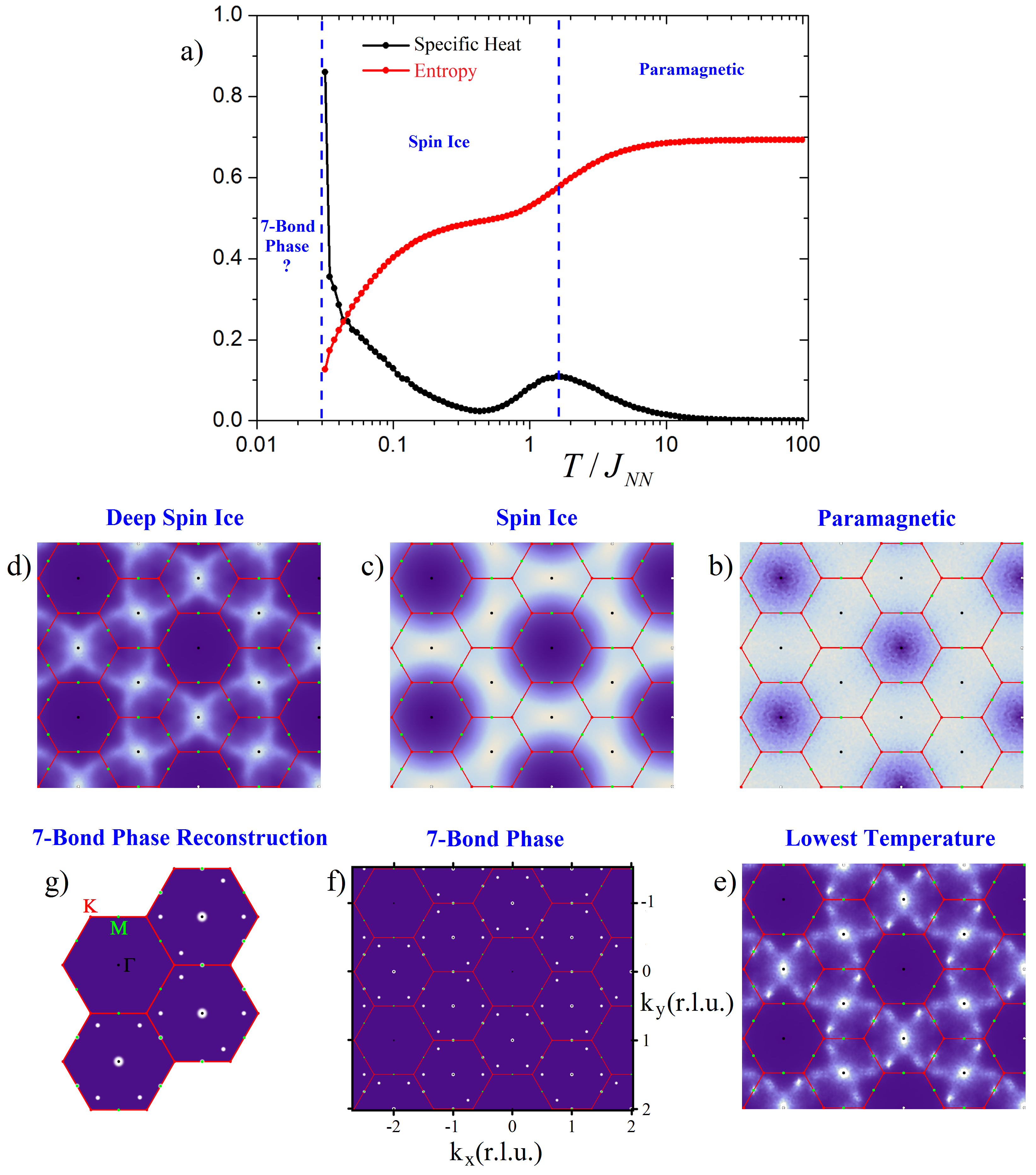}
	\caption{\label{Figure4}
		(Color online) (a) The specific heat and the entropy of dipolar kagome spin ice with $36 \times 36 \times 3$ lattice sites, as sampled by a single-spin flip Metropolis dynamics. The sharp increase in the specific heat signals the presence of an apparent phase transition at $T/J_{NN} \cong 0.03$, deep within the spin ice phase. (b-e) Typical spin structure factor maps are given for each temperature regime and show how the system evolves towards (f) the sixfold-averaged map of the 7-shaped phase, the elementary brick of which is reproduced in (g). Notice how (e) the map of the last sampling temperature still displays a diffuse kagome-like structure typical to (d) a deep spin ice regime, but starts building up the Bragg peaks specific to (f) the 7-shaped phase. This is due to the presence of relatively large clusters of the candidate configuration.}
\end{figure*}

In addition, snapshots sampled in heat-bath conditions at low temperatures contain patches of the candidate configuration. The 7-shaped unit cells can be revealed by drawing an arrow from each spin-up state to all neighboring spin-down states (see Figure~\ref{Figure2}). This so-called arrow picture unveils a tessellation of trapezoidal shapes which seem to struggle to align with the already-established 7-shaped clusters. The presence and distribution of these domains can be quantified for each temperature by computing the spin structure factor, i.e. the Fourier transform of the correlated spin statistics:
\begin{align}
S(\vec{k}) = \sum_{(u,v)} \langle \sigma_u \cdot \sigma_v \rangle \cdot exp[-i \vec{r}_{uv} \vec{k}].
\label{eq:eq2}
\end{align}
where $\sigma_u$ and $\sigma_v$ stand for lattice Ising spins, $r_{uv}$ is their relative position in direct space and $\vec{k}$ is a vector of the reciprocal space.

Figure~\ref{Figure4} presents the temperature plots of the specific heat and the entropy of the dipolar kagome Ising network along with typical spin structure factor maps for each regime. As the system cools and the spins correlate, the diffuse background of the structure factor gradually disappears and Bragg peaks start to develop around certain $\Gamma$ and M points. In addition, there are some faint peaks forming along the $\Gamma$-K lines, at $3/4$ of the distance from the center of the Brillouin zones. In fact, this entire behavior is compatible with the 7-shaped phase (see the maps of Figure~\ref{Figure4}). Its spin structure factor selectively displays Bragg peaks for certain $\Gamma$ and M points, as well as along some $\Gamma$-K lines, at $3/4$ of the distance. This behavior can be understood by referring to the magnetic charge stripe organization. These stripes repeat themselves every two lattice parameters along the $\vec{a_2}$ direction and, along a stripe's direction, display a zig-zag periodicity defined by $\vec{A_1}$, a typical $\Gamma$-K vector (see Figure~\ref{Figure2}). However, given the sixfold degeneracy, all possible domain orientations are present at the last-sampling temperature and the structure factor map is a linear combination of the three possibilities. This suggests, yet again, the compatibility of the proposed ground-state with the ordering tendencies of the dipolar kagome Ising antiferromagnet.

\section{IV. Discussion}  

Although these criteria are necessary but not sufficient conditions for confirming the validity of this configuration as the dipolar ground-state, the 7-shaped phase has passed all these initial tests and stands as a very likely candidate. Interestingly, it displays a 6-fold degeneracy, similar to the ground-states of other kagome-based dipolar spin models \cite{maksymenko_classical_2015,holden_monte_carlo_2015}. 

\begin{figure}
	\includegraphics[width=0.85\linewidth]{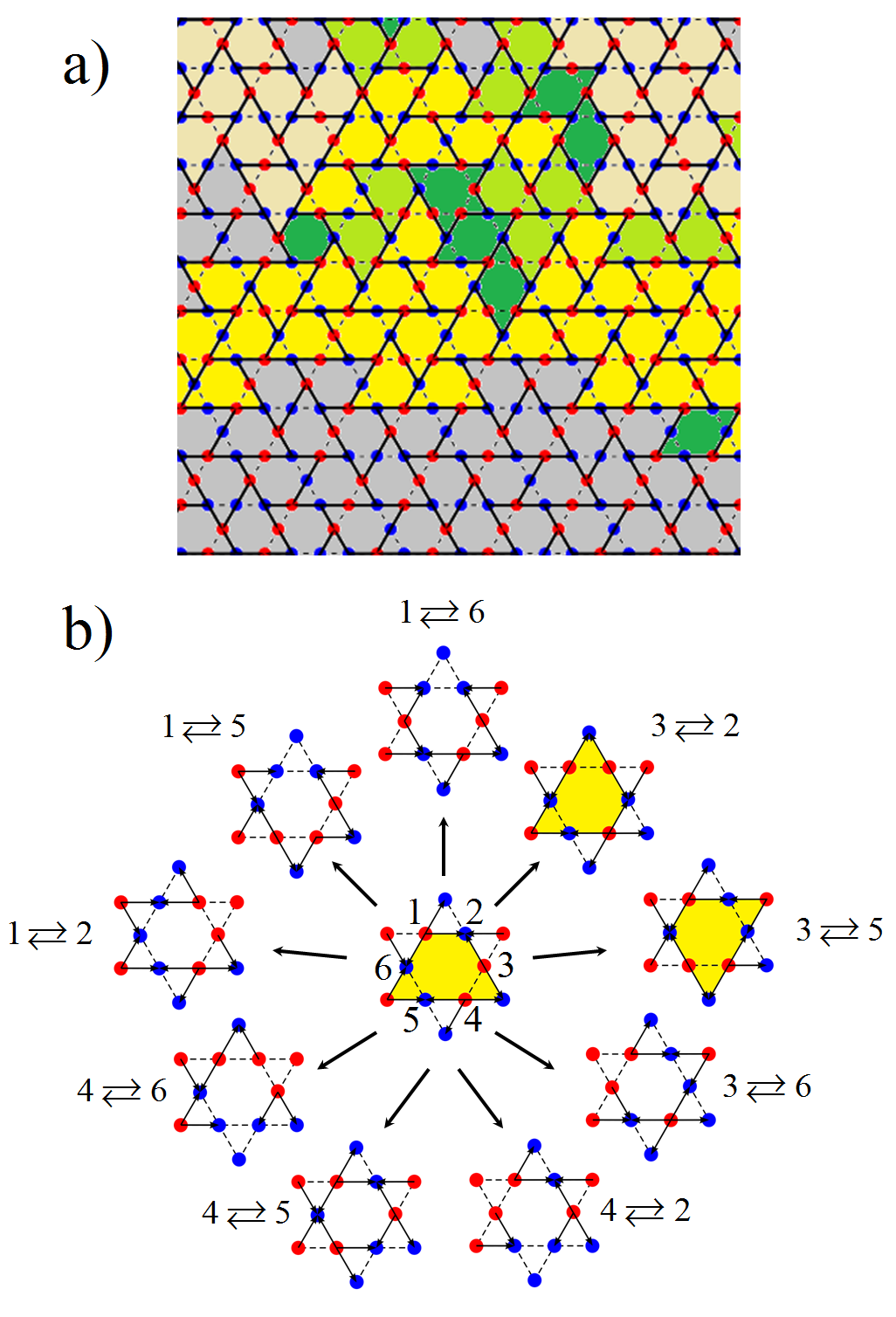}
	\caption{\label{Figure5}
		(Color online) (a) A low-temperature snapshot ($T/J_{NN} \cong 0.03$) from a $36 \times 36 \times 3$ dipolar kagome Ising network showing the formation of 7-shaped domains with different orientation axes (gray, yellow and faded yellow areas). The green areas are trapezoids that do not fit within any of the clusters, while the dark green areas mark defects that do not form preferential trapezoidal structures. Expanding the clusters or adjusting one to fit a neighbor is clearly beyond the practical use of the single spin flip dynamics. However, by interchanging a pair of opposite spins from its inner hexagon (b), the trapezoid's orientation can be modified. Notice that, out of the 9 possible scenarios, only two preserve the initial shape, forming potentially favorable configurations, while other choices can even lead to the creation of kagome ice rule breaking triangles. Therefore, selectively choosing the spin pairs and extending the procedure over larger areas can potentially result in an efficient cluster dynamics that can enhance the exploration of these low-energy manifolds.} 
\end{figure}

Clearly the 12-spin unit cell can be defined in different ways, but the arrow picture that uncovers the 7-shaped sectors presents an intriguing feature. Its arrows are, in fact, projections of the nearest-neighbor stray field lines in the network's plane and reveal how this long-range ordered state translates into a puzzle of short-range favorable spin configurations. While a kagome hexagon with alternating spins would seem to be the building-block ground-state, it is not a preferential formation at the network scale as it cannot properly tile the entire lattice space. Furthermore, through its trapezoidal formations, this arrow picture may prove useful for unraveling an appropriate spin dynamics that can overcome the critical slowing down of single spin-flips, which are rather inefficient in expanding or merging already-formed 7-shaped clusters (see Figure~\ref{Figure5}). By flipping certain pairs of \textit{up} and \textit{down} spins within the internal hexagon of a trapezoidal formation, the trapezoid's orientation can be changed, potentially adjusting it to its environment. This procedure can then be repeated for numerous trapezoids or applied selectively to several at a time in the form of a collective spin dynamics \cite{wolff_collective_1989}.

Another interesting feature of the 7-shaped phase, and of the dipolar kagome Ising antiferromagnet in general, is the formation of magnetic charged stripes. These structures, along with the positive values of the nearest-neighbor charge correlator, highlight the preference for a ferromagnetic charge order, in sharp contrast with dipolar kagome spin ice. However, a perfect ferromagnetic charge crystal might be impossible to achieve at the network scale due to the requirement for global charge neutrality, thus leaving the magnetostatically-favorable option of forming charged lines that effectively screen each-other by alternating. This behavior emphasizes the strong impact that the change of the spin's geometry has on the organization of the emerging charge field.  

\section{V. Conclusion}  

In conclusion, the development of long-range ordered states in the dipolar kagome Ising antiferromagnet have been studied via Monte Carlo simulations, in the quest for the ground-state configuration. In spite of the algorithmic limitations, we can provide a candidate for this previously unknown state through a phenomenological approach. This so-called 7-shaped phase is a rectangular crystal of a 12-spin unit cell which can be revealed in real space through the use of a stray-field based geometrical construction. In reciprocal space, the spin structure factor maps indicate the presence of an alternating zig-zag stripe structure, which is straightforwardly highlighted by the dumbbell charge picture. This configuration passes several test criteria and is compatible with all thermodynamic features observed so far, making it a very likely choice for the dipolar long-range ground state of the kagome Ising antiferromagnet.

From a numerical point of view, the challenge still remains to find a collective spin-dynamics that can properly access this state and even characterize the apparent phase transition. Experimentally, such intriguing spin textures can potentially be achieved within the framework of artificial spin ices, particularly using thermally-active structures, which, for certain protocols, can somewhat overcome the intrinsic slowing down of single-spin flips and locally retrieve low-energy states that are hardly accessible otherwise \cite{chioar_kinetic_2014}.

This work was partially supported by the Agence Nationale de la Recherche through the project ANR12-BS04-009 'Frustrated'. I.A. Chioar acknowledges financial support from the Laboratoire d'excellence LANEF Grenoble and is grateful to Karim Ferhat for several fruitful discussions.


\begin{thebibliography}{}
\bibitem{Note_Cut_off}
{Cutting off the summation over the dipolar couplings can severely affect the system's behavior, particularly in the low-temperature regime. However, for two-dimensional lattices, the dipolar summation is absolutely convergent and, if sufficient terms are considered, the overall thermodynamic features are well captured. We have performed Monte Carlo simulations, as described in the main text, for kagome networks with $L \times L \times 3$ sites, where $L$ can be either $12$, $18$, $21$ or $36$. The same thermodynamic behavior is found in each case, with a slight variation in the position of the specific heat peak that signals an apparent phase transition deep within the spin ice phase. Nevertheless, the simulation is halted at this stage as the driving spin dynamics can no longer ensure the ergodic exploration of these low-energy manifolds.}
\bibitem{Note_Modified_Monte_Carlo_Steps}
{While a Monte Carlo step is genuinely considered as a sweep through the entire lattice, it is not necessarily equal to a unit of Monte Carlo time, which would ensure minimal decorrelation between successively sampled configurations. Therefore, several sweeps are performed between consecutively sampled states and the number of intermediate Monte Carlo steps, called a modified Monte Carlo step, is taken as a multiple of the inverse spin acceptance ratio, which in turn is computed during every thermalization stage.}
\end{thebibliography}
\end{document}